\documentclass[twocolumn,superscriptaddress,preprintnumbers,amsmath,amssymb,aps,prx,longbibliography]{revtex4-1}
\usepackage{graphicx}
\usepackage{dcolumn,xcolor}
\usepackage{amsmath} 
\usepackage{amssymb}
\usepackage{amsfonts}
\usepackage{dsfont}
\usepackage{bm}
\usepackage[latin1]{inputenc}
\usepackage{wasysym}
\usepackage {oplotsymbl}
\usepackage{hyperref} 

\usepackage[normalem]{ulem}

\definecolor{myyellow}{rgb}{1, 0.8, 0.2}
\definecolor{mygreen}{rgb}{0.6, 0.8, 0.3}

\begin{document}
\title{Creep in reactive colloidal gels: a nanomechanical study of cement hydrates}
 
\author{Michael Haist}
\email{michael.haist@baustoff.uni-hannover.de}
\affiliation{Leibniz University Hannover, Institute for Building Materials, Germany}
\affiliation{Massachusetts Institute of Technology, Department of Civil and Environmental Engineering,\\ 77 Massachusetts Avenue, Cambridge, Massachusetts 02139, USA}
\author{Thibaut Divoux}
\email{thibaut.divoux@ens-lyon.fr}
 \affiliation{MultiScale Material Science for Energy and Environment, UMI 3466, CNRS-MIT, 77 Massachusetts Avenue, Cambridge, Massachusetts 02139, USA}
 \affiliation{University of Lyon, Ens de Lyon, Univ Claude Bernard, CNRS, Laboratoire de Physique, F-69342 Lyon, France\looseness=-1}
 \author{Konrad J. Krakowiak}
 \affiliation{University of Houston, Civil and Environmental Engineering Department, Cullen College of Engineering, 4726 Calhoun Road, Houston, TX 77204-4003, USA}
 \author{J{\o}rgen~Skibsted}
 \affiliation{Aarhus University, Laboratory for Solid-State NMR Spectroscopy of Inorganic Materials, Interdisciplinary Nanoscience Center (iNANO) and Department of Chemistry, Denmark}
 \author{Roland J.-M. Pellenq}
 \affiliation{MultiScale Material Science for Energy and Environment, UMI 3466, CNRS-MIT, 77 Massachusetts Avenue, Cambridge, Massachusetts 02139, USA}
\author{Harald S. M{\"u}ller}
 \affiliation{Karlsruhe Institute of Technology, Institute for Concrete Structures and Building Materials, Karlsruhe, Germany}
 \author{Franz-Josef Ulm}
 \affiliation{Massachusetts Institute of Technology, Department of Civil and Environmental Engineering,\\ 77 Massachusetts Avenue, Cambridge, Massachusetts 02139, USA}

\date{\today}

\begin{abstract}
From soft polymeric gels to hardened cement paste, amorphous solids under constant load exhibit a pronounced time-dependent deformation called creep. The microscopic mechanism of such a phenomenon is poorly understood in amorphous materials, and constitutes an even greater challenge in densely packed and chemically reactive granular systems. Both features are prominently present in hydrating cement pastes composed of calcium silicate hydrate (C-S-H) nanoparticles, whose packing density increases as a function of time, while cement hydration is taking place. Performing nano-indentation tests and porosity measurements on a large collection of samples at various stages of hydration, we show that the creep response of hydrating cement paste is mainly controlled by the inter-particle distance and results from slippage between (C-S-H) nanoparticles. Our findings provide a unique insight into the microscopic mechanism underpinning the creep response in aging granular materials, thus paving the way for the design of concrete with improved creep resistance. 
\end{abstract}

\maketitle
\section{Introduction}
Under constant external load, solid materials display an instantaneous elastic strain followed by pronounced time-dependent deformations. Such a mechanical response, called creep, is ubiquitously observed in both crystalline (i.e., ordered) and amorphous (i.e., disordered) materials, such as gels, glasses, granular systems and hardened cement paste \cite{Weibull:1939,Nguyen:2011,Vandamme:2009}. While the creep behavior of crystalline materials has been quantitatively associated to the collective motion of dislocations in such materials \cite{Miguel:2002,Jakobsen:2006,Csikor:2007,Miguel:2008,Vinogradov:2012}, the microscopic mechanism underpinning the creep response in amorphous granular materials remains controversial and a topic of intense research \cite{Sentjabrskaja:2015,Cao:2017,Aime:2018,Liu:2018,Cipelletti:2020}.

Indeed, in granular matter the creep response is highly sensitive to the particle volume fraction as well as the inter-particle interactions. On the one hand, in colloidal gels, i.e., systems composed of attractive particles forming a percolated network at low volume fraction, the early stage creep response is often accounted for by linear viscoelastic properties, i.e., macroscopically reversible deformations, up to a critical strain value beyond which a sudden burst of localized plastic rearrangements triggers the nucleation of microscopic cracks long before the macroscopic failure of the material \cite{Siebenburger:2012a,Leocmach:2014,Aime:2018}. 
On the other hand, in densely packed granular materials, creep is often ascribed to a disturbance of the force network present in the material microstructure, which corresponds to plastic rearrangements \cite{Kuhn:1993,Bowman:2003}. In the limit of jammed systems, which lack sufficient free volume for local reorganization of the granular network, the detailed mechanisms of creep are not fully understood. Creep here is often associated with a dilation of the particle system \cite{Reynolds:1885,Behringer:2019}, and for large stresses, the individual deformation of the particles and their fracture has been observed to strongly enhance the creep process \cite{McDowell:2003,Shi:2019}. 

Identifying and modelling the microscopic mechanism of the creep response in amorphous granular materials is further complicated by the presence of chemical reactions such as dissolution, precipitation and chemical bonding, which result in irreversible aging and a pronounced morphological evolution of the materials microstructure \cite{Raj:1981,Meer:1997,Spiers:1990,Dysthe:2002}. The interplay between the changes induced by these reactions and those triggered by the external loading make the creep response at the microscopic scale much more complex to decipher. This is the case for hardened cement paste, which is the ``glue" that provides strength to concrete \cite{Mueller:2009}. The reactive nature of this system and its time-evolving microstructure pose a significant challenge for studying its creep behaviour, and at the same time offer a unique opportunity for determining the impact of packing density on the creep response of granular materials in general. 

The mechanical properties of hardened cement paste are governed by colloidal particles of Calcium-Silicate-Hydrate (C-S-H), which precipitate as nanoparticles after mixing and dissolution of the polydisperse cement powder in water \cite{Jennings:2000,Bullard:2011}. These colloidal particles interact via short-range attraction and longer-range electrostatic repulsion forces, which are strongly influenced by the ion content and composition of the solvent (with ions resulting from the cement dissolution) \cite{Pellenq:1997,Pellenq:2004,Plassard:2005,Pellenq:2008,Ioannidou:2016}. Eventually, hardened cement paste forms a complex nanoporous network of colloidal particles, which is age and composition dependent. In consequence, depending on the degree of hydration $\xi$ (defined as the fraction of cement powder having reacted with water at a certain point in time), the water-to-cement mass ratio $w/c$, the age of the sample when the mechanical load is applied, and the moisture conditions, hardened cement paste shows a broad range of creep responses under a constant external load  \cite{Wittmann:1973,Parrott:1973,Tamtsia:2000,Powers:1968}. 

\begin{figure*}[!th]
\centering
\includegraphics[width=0.7\linewidth]{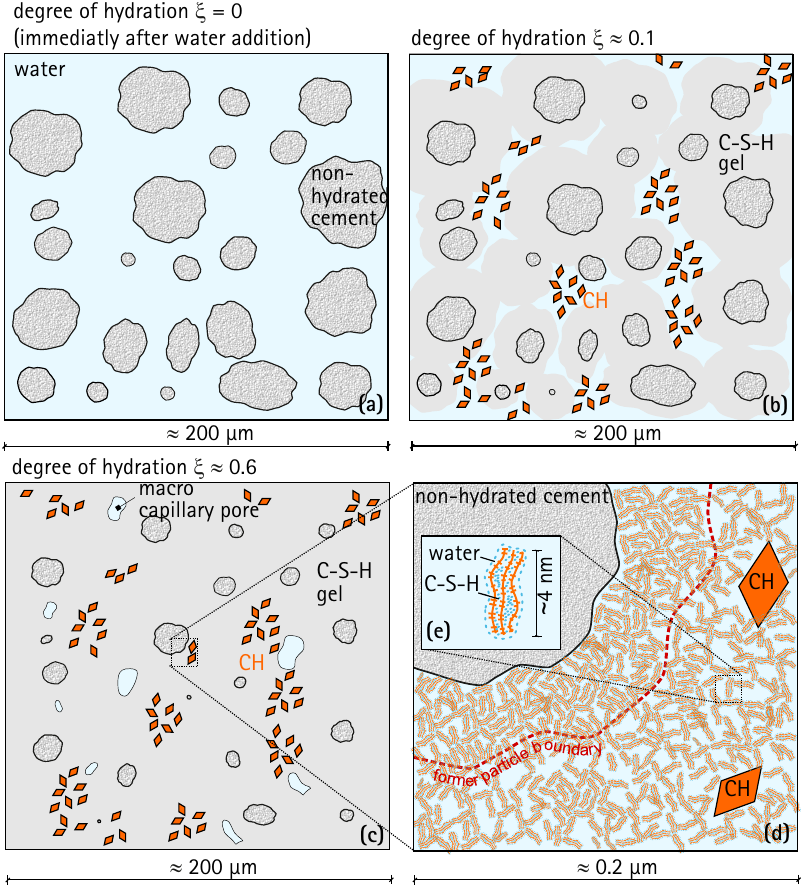}
\caption{(color online) Sketch depicting the hydration process of cement particles in water (a), which results in the formation of hardened cement paste consisting of non-hydrated cement particles, C-S-H phases, CH phases, and pores of different size and content (b and c); due to the hydration process cement particles dissolve and C-S-H and CH precipitate in the solvent in a less densely packed manner outside the (former) particle surface and in a tighter manner within the bounds of the former particle surface (which was dissolved) (d); gel pores designate voids between the C-S-H particles; capillary pores designate voids formerly occupied by the solvent \label{fig0}}
\end{figure*}

To date, three different microscopic mechanisms have been proposed to account for the macroscopic reversible and irreversible creep response of hardened cement paste. The underlying concepts however can be generalized and adapted to basically any kind of granular system. These scenarios explain the creep behavior either as a shear deformation of the individual C-S-H colloidal particles (length scale approx. 4 nm) \cite{Feldman:1972}, as a slippage between these particles (length scale approx. up to 20 nm) \cite{Powers:1968}, or as the formation of microcracks at spatial scales significantly larger than the particle size \cite{Wittmann:1982}. Despite substantial evidence for the existence of all three of these mechanisms \cite{Brown:1976,Bentur:1979,Bazant:1989,Bazant:1997}, their respective contribution to the macroscopic creep response is unknown. Moreover, these static scenarios neither take into account the reactive and time-dependent nature of cement paste, nor its heterogeneous microstructure. Indeed, hardened cement paste is not homogeneous on a microstructural level, but is primarily formed by the precipitation of C-S-H particles, which however prevail in regions of low and high (packing) density. These regions are  often referred to as LD-C-S-H and HD-C-S-H, respectively \cite{Jennings:1981,Diamond:1993,Ioannidou:2016b}. Their effect onto the creep behaviour and creep mechanisms so-far has not been considered. Finally, insight into the different creep mechanisms on the sub-micron and nanoscale was only recently gained using numerical simulations \cite{Manzano:2013,Morshedifard:2018}. Whilst ref.~\cite{Manzano:2013} reports clear evidence for creep deformation of individual particles, ref.~\cite{Morshedifard:2018} showed that the main contributing factor to creep must be seen in a slippage between particles. Unfortunately, substantial experimental evidence to test either of these numerical results is still missing.

Building upon the statistical analysis of an extensive series of nano-indentation tests on cement paste samples having reached different degrees of hydration, we provide a local scenario for the creep response of hardened cement paste that takes into account the heterogeneous and time-evolving properties of cement's microstructure. Spatially-resolved mechanical properties are coupled to elemental chemistry surface mapping to isolate the mechanical properties of the so-called LD- and HD-C-S-H at the sub-micron level. As a key result, we show that for both types of C-S-H phases, the major contribution to the creep response is set by the slippage between C-S-H nanoparticles. 
Our results show that LD- and HD-C-S-H phases only differ by their respective volume fraction in C-S-H particles, and thus behave as a single granular phase when it comes to the creep response. 
In that framework, the creep response of hardened cement paste is mainly controlled by the inter-particle distance, which decreases for increasing degree of hydration as more C-S-H particles precipitate. Finally, we further confirm this picture by porosimetry measurements, which allow us to quantify the porosity of the cement paste as a function of the degree of hydration. The obtained porosity information further is essential, to connect the findings on hardened cement paste to the creep behaviour of densely packed and possibly jammed granular systems in general. 

The outline of the manuscript goes as follows: Section~\ref{overview} provides an overview regarding cement hydration in terms of chemical reaction, and describes the emergence of hardened cement paste microtexture. Section~\ref{mechanics} recalls the state-of-the-art regarding key factors controlling the mechanical properties of hardened cement paste. We then present our experimental results in section~\ref{results}, where we report the sub-micron creep response of the paste as a function of the packing density of the C-S-H particles, as well as microstructural measurements by NMR spectroscopy and porosimetry. Finally, we summarize our main conclusions in section~\ref{DiscussionConclusion}.

\section{Overview on cement hydration}
\label{overview}
Hardened cement paste (HCP) forms during the reaction of Ordinary Portland Cement -- primarily consisting of Calcium Silicate phases 3CaO$\cdot$SiO$_{2}$  and 2CaO$\cdot$SiO$_{2}$ (abbreviated C$_{3}$S and C$_{2}$S, respectively) and minor fractions of Calcium Aluminate phase 3CaO$\cdot$Al$_{2}$O$_{3}$ (C$_{3}$A) and Ca$[$SO$_{4}]\cdot$2H$_{2}$O -- with water. In the rest of the manuscript, the oxide phases in cement chemistry are abbreviated as follows: C for CaO, S for SiO$_{2}$, H for H$_{2}$O, and A for Al$_{2}$O$_{3}$.

From the hydration of calcium silicates, two products form via a dissolution-precipitation reaction: amorphous calcium-silicate-hydrate (C-S-H) and crystalline calcium-hydroxide (CH), with the amount of C-S-H being formed clearly outweighing the CH phase \cite{Bullard:2011}. Moreover, C-S-H forms via precipitation at a density of about 2.45 g/cm$^3$, a mean CaO$/$SiO$_{2}$ (or C/S)-ratio of 1.7, forming brick-like particles with a typical edge length of approx.~4.2~nm [Fig.~\ref{fig0}(e)] 
that show a layered microstructure consisting of calcium-silicate-hydrate and water  \cite{Jennings:2000,Jennings:2008}. 
These particles randomly aggregate to form larger structures, incorporating more or less water, thus showing significant differences in packing density $\eta$ and density as sketched in Fig.~\ref{fig0}(d). The water incorporated inside  as well as between the particles is referred to as \textit{gel water}, which occupies the so-called gel porosity, with pores of up to 30~nm in diameter [Figs.~\ref{fig0}(d) and \ref{fig0}(e)]. 

Whilst CH primarily precipitates in the bulk of the solvent as plate-like crystals with a typical size reaching up to several microns [Fig.~\ref{fig0}(a-d)], C-S-H particles nucleate on the surface of the dissolving cement particle, thus creating a shell of hydration products around the latter, and in turn, increasingly hinder the dissolution process of the cement [Fig.~\ref{fig0}(d)] \cite{Tennis:2000}. Due to this shell-like growth mechanism, seminal works on cement hydration started differentiating between two types of C-S-H, namely, ($i$) C-S-H forming outside of this shell in the free space between cement particles yielding a loose-packing assembly of C-S-H particles, thus termed outer product or low-density C-S-H (LD-C-S-H), and ($ii$) C-S-H forming within the shell and thus the space cleared by the progressing dissolution of the cement, showing a significantly higher packing density and thus being termed inner product or high-density C-S-H (HD-C-S-H) \cite{Jennings:1981}. LD- and HD-C-S-H are chemically quasi identical with respect to their solid constituents. However possess a different packing density, porosity and density upon they can be clearly distinguished. The packing-density distributions of LD- and HD-C-S-H show a pronounced overlap. 

Originally it was believed by Jennings and others (see e.g. \cite{Jennings:2000}), that the mentioned properties such as packing density are imminent to the phases, however in this research we can clearly show that they evolve in time as a function of hydration.   
Both LD- and HD-C-S-H as well as CH (in the following termed as cement gel) form in the space between the cement particles [Fig.~\ref{fig0}(a)-(d)]. By chemically and physically binding water, the volume of C-S-H and CH formed during this hydration process is approx.~2.13 times greater than the volume of the hydrated cement \cite{Powers:1948}. In terms of microtexture, the hydration process thus consists in filling up the space between the cement particles (originally filled with water) with a mix of C-S-H gel (more or less densely packed C-S-H particles) and CH. 
In the early stage of hydration, hardened cement paste is highly porous, and the corresponding voids are referred to as \textit{capillary pores}, with a diameter up to several microns. With continuing hydration, the capillary pore space is filled with C-S-H gel and CH, leading to a densification of the system. For pastes with high water content (and thus relatively low cement content), the amount of gel formed during hydration does not suffice to fill the entire space. In practice, for a water to cement mass ratio smaller than 0.4, large capillary pores remain in the paste even when the entire cement powder has reacted \cite{Powers:1948}. In conclusion,  gel pores form due to the chemical reaction process, whereas capillary pores are remnants of the interstitial space between the original cement particles. 

\begin{figure*}[!t]
\centering
\includegraphics[width=\linewidth]{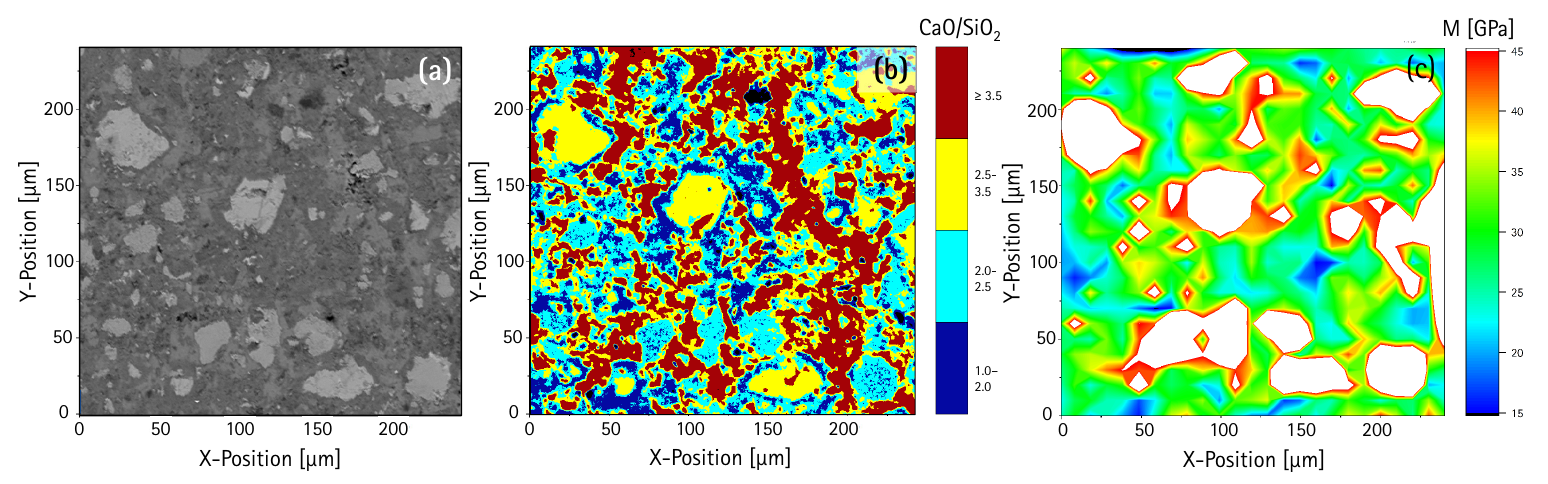}
\caption{(color online) (a) Scanning electron microscope (SEM) backscattered image of the surface of a sample of hardened cement paste. Spatial distribution of (b) Map of the CaO/SiO$_{2}$-ratio determined via WDS element mapping of Ca and Si and converted into oxide ratio, and (c) map of the indentation modulus $M$ determined by nanoindentation. Data measured on a cement sample with a degree of hydration $\xi_{\rm NMR}=0.6$. 
\label{fig1a}}
\end{figure*}

\section{Mechanical properties of hardened cement paste} \label{mechanics}

The macroscopic mechanical properties of HCP are primarily governed by the capillary porosity and the degree of hydration $\xi$, i.e., the share of cement, which has already dissolved to form C-S-H gel and CH. With increasing $\xi$ and thus decreasing capillary porosity, the strength of HCP increases, and the creep rate (i.e., the change in creep deformation over time) decreases \cite{Powers:1948,Vandamme:2013}. 
Nonetheless, even a fully hydrated HCP with $\xi=1.0$ with no capillary pores (i.e., a water-to-cement ratio w/c lower than 0.4) shows pronounced creep. Thus a closer look has to be given to the mechanical properties of the gels constituents and their corresponding mechanical interaction behaviour.
Moreover, HCP is a highly porous system with pore sizes ranging from fractions of a nanometer to microns; hence the mechanical properties at all scale levels are also strongly influenced by the water content of the system. Gel pores are normally fully filled with water, whereas water in the capillary pores is predominantly present in the form of water films \cite{Powers:1948,Young:1982,Zhou:2019}. In consequence, depending on the degree of hydration, the water to cement mass ratio w/c and the moisture conditions, HCP shows only minor creep for dry conditions at very low w/c-ratios and very high degree of hydration and very substantial creep in the opposite case when subjected to a constant external load \cite{Wittmann:1973,Parrott:1973,Tamtsia:2000}.

Concerning the sub-micron mechanical properties of the cement gel, molecular dynamics simulations on individual C-S-H particles have shown that clear trends between the Ca/Si-ratio of the particles, and the sub-micron mechanical properties can be established \cite{Abdolhosseini:2014}. For a typical Ca/Si ratio as observed in C-S-H gel from Portland cement hydration, the stiffness M of the individual particles was determined to be about 72~GPa. 
Investigations regarding the interaction of the individual C-S-H particles revealed that the latter interact via short-range attraction and longer-range electrostatic repulsion forces, which are strongly influenced by the ion content and composition of the pore water \cite{Plassard:2005}. Furthermore, significant efforts in simulating these colloidal interactions allowed for an accurate modeling of the C-S-H gels' physical (and mechanical) properties \cite{Pellenq:1997,Pellenq:2004,Pellenq:2008,Ioannidou:2016,Morshedifard:2018}. In particular, these simulations closely reproduce results obtained from experimental nanoindentation on HCP \cite{Ioannidou:2016b,Abdolhosseini:2014}.

\section{Results}
\label{results}

In order to determine the creep response of hardened cement paste at different stages of the hydration process, we have prepared a series of 120 samples, whose hydration process is stopped at 30 different points in time by solvent exchange with isopropanol (see Appendix~A for details) \cite{Zhang:2011}. These samples were subjected to various investigation techniques such as statistical nanoindentation (see Appendix~B), chemical surface mapping (see Appendix~C) and various bulk measurement techniques (see Appendix~D). The sample preparation process, conducted over a year, yielded a series of samples with various hydration degrees $\xi_{\rm NMR}$ for the cement silicate phases (i.e., Alite and Belite) ranging from 0.2 to 0.7 as determined by \textsuperscript{29}Si NMR spectroscopy (see Appendix~D for details). Note that this NMR method is primarily focussed on the hydration behaviour of the calcium-silicate phases and neglects the contribution of calcium-aluminate phases to hydration. To monitor the evolution of the cement paste's mechanical properties at the microscale for increasing degree of hydration, we have conducted nanoindentation tests to determine the hardness $H$, the indentation modulus $M$, and the creep modulus $C$ at a scale of approximately 1~$\mu$m$^3$. 

The indentation hardness $H$ is defined as the ratio of the maximum force $P_{\rm max}$ applied by an indenter onto the sample, and the projected contact area $A_{c}$ between the intender and the indented surface \cite{Oliver:1992}. The hardness $H$ correlates to the strength of the sample. The indentation modulus $M$ relates to the elasticity of the indented material, and is defined as $M=E/(1-\nu^2)$, where $E$ is the Young Modulus, and $\nu$ the Poisson's ratio of the sample. In practice, $M$ is determined following the method of Oliver and Phaar by measuring the slope $S=dP/dh|_{h=h_{\rm max}}$ during the unloading step of an indentation experiment, and computing $M=S/(2a_{u})$, where  $a_{u}=\sqrt{A_{c}\pi}$ is the radius of contact between the indenter probe and the indented surface upon unloading \cite{Oliver:1992,Oliver:2004}. Finally, the indentation creep modulus $C$ is derived from the contact creep compliance $L(t)$, which verifies $L(t)=2a_{u} \Delta h(t)/P_{\rm max} +const$, for short experiments, i.e., limited to a few minutes \cite{Vandamme:2012}. For a logarithmic time-dependence of the creep phenomenon, i.e., $\Delta h(t) \sim x_1 \ln(1+t/\tau)$ as observed here, the rate of creep compliance reads $dL(t)/dt = 1/(Ct)$, with $C=P_{\rm max}/(2a_ux_1)$. In that framework, the parameter $C$ has the unit of a stress as well, but, as the load $P_{\rm max}$ is constant in the previous equation, $C$ is quasi proportional to the inverse creep rate, i.e., the (inverse) velocity of the creep process. For more details on the experimental determination of $H$, $M$ and $C$, see Appendix~B. 

In the following, we restrain the data analysis to continuous force-depth data, whereas outliers including curves with discontinuities such as jumps, etc.~are not considered (see Section~\ref{crack} for the analysis of force-depth data with jumps).
A sample is typically characterized by a nanoindentation grid of 25$\times$25 indentations yielding 625 triplets ($H$, $M$, $C$), that are statistically analyzed with a Gaussian Mixture Modelling approach (see Appendix~B for details and ref.~\cite{Krakowiak:2015}). This procedure allowed us to identify up to 5 different phases, i.e., two types of C-S-H-phases, a CH-phase, non-reacted cement particles, and a mix-phase consisting of C-S-H and CH \cite{note1}. 

We illustrate the relevance of the clustering method to capture the properties of the individual phases composing HCP on a sample with a degree of hydration $\xi_{\rm NMR} = 0.6$. A picture obtained by electronic microscopy of a region of interest on the sample surface is shown in Fig.~\ref{fig1a}(a). The corresponding spatial distribution of the CaO/SiO$_{2}$-ratio  obtained via elemental chemical mapping is shown in Fig.~\ref{fig1a}(b) (for details on the elemental chemical mapping, see Appendix~C), while that of the indentation modulus $M$ obtained by nanoindentation is shown in Fig.~\ref{fig1a}(c). 
The non-hydrated cement grains visible as light gray areas on the SEM image are clearly visible in both the chemical and mechanical maps. These grains correspond to the areas with the highest value of $M$ in Fig.~\ref{fig1a}(c) outweighing  that of the hydration product surrounding the particles and to a CaO/SiO$_{2}$-ratio ranging between 2.0 and 3.5 in Fig.~\ref{fig1a}(b), reflecting the dominant effect of the mineral tricalcium-silicate (CaO/SiO$_{2}$ = 3.0) making up the cement. 
Some remnants of cement grains consisting purely of dicalcium-silicate with CaO/SiO$_{2}=2.0$ are also visible in Fig.~\ref{fig1a}(b), see e.g., coordinates $X=150~\mu$m, $Y =50~\mu$m), and their indentation modulus also significantly exceeds that of the reaction products Fig.~\ref{fig1a}(c). Due to their significantly higher indentation modulus $M$ (and hardness $H$; not shown) the cement particles are correctly identified by the Gaussian Mixture algorithm as individual clusters.

The map corresponding to the CaO/SiO$_{2}$-ratio further shows clusters with very high CaO/SiO$_{2}$-ratios [see red color in Fig.~\ref{fig1a}(b)], hinting to patches dominated by CH. These patches display a channel-like distribution, reflecting their formation history, i.e., CH precipitating in the free water between the cement particles. Looking at the mechanical map [Fig.~\ref{fig1a}(c)] the CH patches only lead to a minor increase in hardness, due to the fact that the indentation modulus of pure CH is insignificantly higher (approx. factor 1.5) than that of C-S-H. The Gaussian Mixture Modelling approach identifies these patches as individual clusters. 
 
 Finally, we observe in Fig.~\ref{fig1a}(b) that the space between the cement particles and the CH channels is filled with a mineral phase with a CaO/SiO$_{2}$-ratio between 1.5 and approx.~2.5, which is identified as C-S-H. The mechanical properties of this phase determined by indentation further allow the distinction of two distinct sub-phases, their mechanical properties corresponding to the mechanical properties of LD- and HD-C-S-H as identified in previous research \cite{Krakowiak:2015}.
 
\begin{figure*}[!th]
\centering
\includegraphics[width=0.8\linewidth]{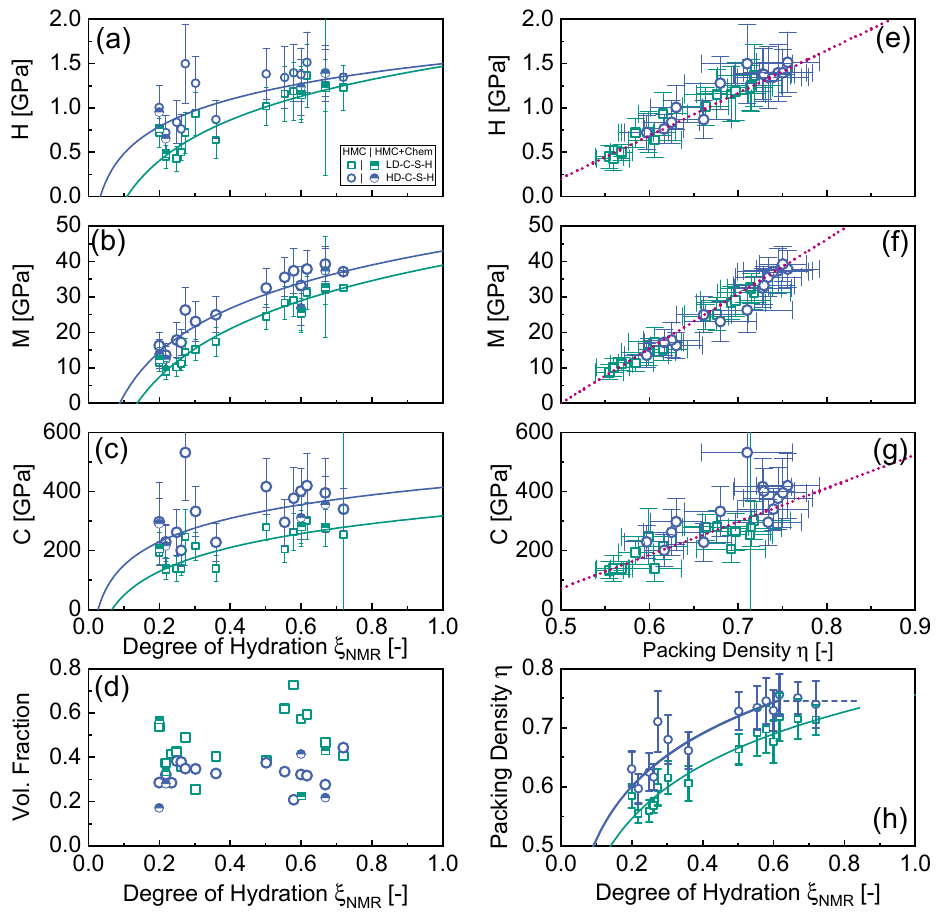}
\caption{(color online) Mechanical properties as a function of the degree of hydration: (a, b, c) indentation hardness $H$, indentation modulus $M$ and creep modulus $C$ for LD-C-S-H and HD-C-S-H phases determined by statistical nanoindentation as a function of degree of hydration $\xi_{\rm NMR}$ determined by \textsuperscript{29}Si NMR spectroscopy. Both phases are identified by a statistical analysis based on $H$, $M$ and $C$ (index HMC, open symbols) or based on $H$, $M$, $C$ and the chemical composition in Calcium, Silicon and Aluminum at the locus of the indents (index HMC-chem, half-filled symbols) as input values. Continuous lines are logarithmic functions, which serve as guidelines for the eye. (e, f, g) same data $H$, $M$ and $C$ as in (a, b, c) vs.~packing density $\eta$ computed by micromechanical modelling (see text) of C-S-H particles in the respective phases. Dashed lines correspond to the best linear fit of the data. (d)~Volume fraction $f$ and (h) packing density $\eta$ of LD- and HD-C-S-H phases vs.~degree of hydration $\xi_{\rm NMR}$. The blue dashed line corresponds to a packing density of 0.74, and continuous lines serve as guide for the eye. In all the graphs, the error bars stands for twice the standard deviation of the considered observable.
\label{fig1}}
\end{figure*}

\subsection{Impact of cement hydration on the sub-micron mechanical response of hardened cement paste} 

Figure~\ref{fig1} shows an overview of the entirety of the mechanical results for the two phases with the lowest mechanical properties as identified by the Gaussian Mixture algorithm. Following the reasoning outlined with Fig.~\ref{fig1a}, these two phases can be classified as LD- and HD-C-S-H, and their mechanical properties are in agreement with that of the literature \cite{Sorelli:2008,Vandamme:2009}. The composition of the samples is dominated by these two phases [Fig.~\ref{fig1}(d)], making up for more than 75\% of the hydration products, with the individual contents, however, being subject to increased scatter. The volume fraction of LD-C-S-H increases with the degree of hydration $\xi_{\rm NMR}$, especially for $\xi_{\rm NMR} \gtrsim 0.5$, whereas the content of HD-C-S-H remains constant at a volume fraction of about 35\% over the entire range $0.2\lesssim \xi_{\rm NMR} \lesssim 1$. Remarkably, both LD- and HD-C-S-H show increasing values of $H$, $M$ and $C$ for increasing degree of hydration $\xi_{\rm NMR}$ [Fig.~\ref{fig1}(a)-(c)]. 
Note that these results are robust, whether the phase determination is obtained by clustering the results solely based on the mechanical properties ($H$, $M$, and $C$) or by additionally taking into account the chemical (elemental) composition in Ca, Si, and Al determined at the location of the indents by Wave Dispersion Spectroscopy [see half-filled symbols in Fig.~\ref{fig1}(a)--(d)]. See also Appendix C for more details. 

The LD- and HD-C-S-H phases can be modeled as an assembly of C-S-H nanoparticles interacting by cohesive forces \cite{Jennings:2000}. Using a micromechanical approach introduced in refs.~\cite{Ulm:2007,Cariou:2008} and assuming the C-S-H particle stiffness to be 72~GPa \cite{Abdolhosseini:2014}, we determine the packing density $\eta$ of each phase as a function of the degree of hydration [Fig.~\ref{fig1}(h)]. This determination was based on the solution of the inverse problem, formulated within the framework of cohesive granular materials, in which a least-square minimization was employed to determine the packing density $\eta$, as well as particles' intrinsic properties \cite{Ulm:2007}.
 For both LD- and HD-C-S-H phases, $\eta$  increases with $\xi_{\rm NMR}$ [Fig.~\ref{fig1}(h)]. The difference in packing density $\Delta \eta$ between the two phases remains constant, equal to approximately 0.05, for degrees of hydration up to $\xi_{\rm NMR} \simeq 0.55$, before decreasing and vanishing in the limit of complete hydration ($\xi_{\rm NMR}=1$). This result supports the picture that the formation of new LD-C-S-H phases in the larger capillary pore space of hydrating cement paste is significantly influenced by the free volume available for this precipitation process to occur, leading to more densely packed phases at higher degrees of hydration. Moreover, the formation of new LD-C-S-H for increasing degrees of hydration goes along with a continuous compaction of the existing LD-C-S-H phases, shifting their classification from LD- to HD-C-S-H together with a compaction of the existing HD-C-S-H phases. The terminal packing density $\eta_{\rm max}$ is limited to values of about 0.75 [Fig.~\ref{fig1}(h)], in good agreement with the maximum packing density of monosized spheres \cite{Jennings:2008}. In that framework, the mechanical properties of LD- and HD-C-S-H phases can now be plotted as a function of the packing density $\eta$ of C-S-H particles [Fig.~\ref{fig1}(e)--(g)]. $H$, $M$ and $C$ grow linearly (at least in the considered range) for increasing packing density and no distinction can be made between the results of the LD- and HD-C-S-H phases. This result is remarkable and shows that the packing of the individual C-S-H nanoparticles is the fundamental parameter determining the mechanical properties of both LD- and HD-C-S-H phases. Moreover, this result suggests there is a continuum between these two phases that only differ by their packing density. 

\begin{figure}[t!]
\centering
\includegraphics[width=\columnwidth]{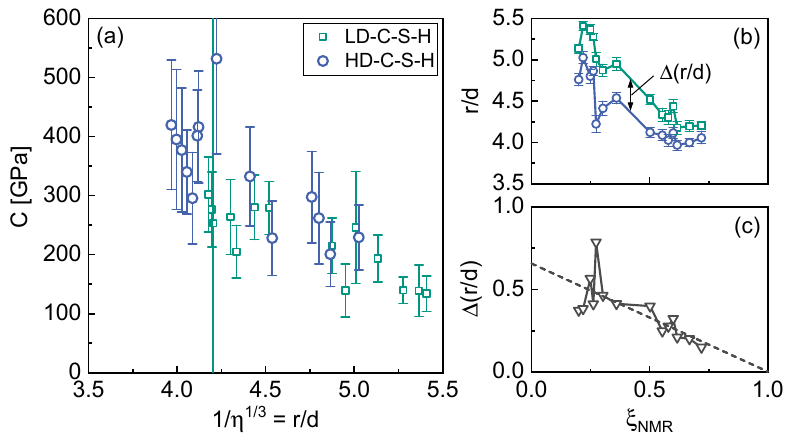}
\caption{(color online) (a) Creep modulus $C$ as function of $r/d$, the ratio of the distance between C-S-H nanoparticles in LD- and HD-C-S-H phases normalized by the nanoparticle size. (b) Interparticle spacing $r/d$ for LD- and HD-C-S-H phases and (c) difference between LD- and HD-C-S-H interparticle distance $\Delta(r/d)$ as function of the degree of hydration $\xi_{\rm NMR}$. The dashed line in (c) corresponds to the best linear fit of the data.
\label{fig2}}
\end{figure} 

Following the approach employed on mesoscale simulations \cite{Ioannidou:2016b}, the packing density $\eta$ can be used to estimate the ratio between the average interparticle distance $r$ and the diameter $d$ of the C-S-H particles as follows: $r/d=\eta^{-1/3}$ \cite{Abuhaikal:2018}. As can be seen in Fig.~\ref{fig2}(a), the creep modulus $C$ increases roughly linearly with decreasing interparticle distance $r/d$, over the range of packing densities that can be accessed experimentally. This trend corresponds to a significant reduction in creep rate and strongly suggests that the creep response of hardened cement paste primarily depends on interparticle slippage processes, and not on the creep behavior of the individual C-S-H particles. Hence, early-age creep occurs at the inter-granular scale and not at a spatial scale smaller than the size of an individual C-S-H particle. Moreover, Fig.~\ref{fig2}(b) shows that this process is valid both for LD- and HD-C-S-H, as both HD- and LD-C-S-H packing fraction distributions exhibit significant overlap. For $\xi_{\rm NMR} \geq$~0.6 the interparticle distance strives to an end value, indicating that the hydration process seems to be controlled by packing limitations for high degrees of hydration. Finally, the difference in interparticle distance between LD- and HD-C-S-H decreases for increasing degree of hydration [Fig.~\ref{fig2}(c)]. Both results indicate that the system is entering a jammed state for high degrees of hydration, which significantly slows down creep -- but does not prevent it totally.

\begin{figure}[t!]
\centering
\includegraphics[width=0.85\linewidth]{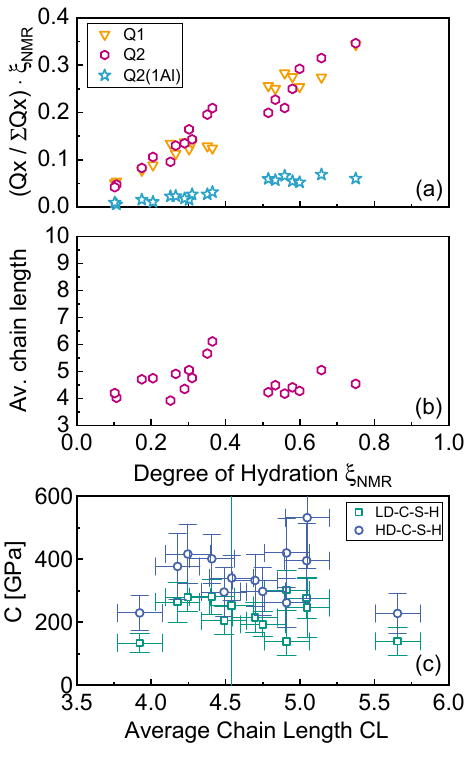}
\caption{(color online) (a) Fraction of C-S-H dimers and silicate end groups (designated $Q^1$), polymeric units of silicate tetrahedra($Q^2$), including sites bound to tehtrahedral Al in the chains ($Q^2(1Al)$) relative to the total amount of C-S-H phases formed multiplied by the degree of hydration $\xi_{\rm NMR}$ and plotted as a function of $\xi_{\rm NMR}$. (b) Average alumino-silicate chain lengths (CL) of the C-S-H polymers (see Appendix~D for calculations) as a function of $\xi_{\rm NMR}$. The average standard deviation is estimated to $\pm$ 0.15 for the chain length values. (c) Creep modulus $C$ vs average alumino-silicate chain length of the C-S-H polymers, for both LD- and HD-C-S-H data.
\label{fig3}}
\end{figure} 

\subsection{Creep response controlled by inter-particle sliding friction} 
Additional support for these findings are derived from $^{29}$Si magic angle spinning (MAS) NMR experiments, which allow us to identify the different types of condensation of silicate tetrahedra in the C-S-H-nanoparticles, i.e., dimers and silicate end groups (denoted $Q^1$), members of silicate chains ($Q^2$) and silicate tetrahedral neighboring $Al$ tetrahedron in the chain ($Q^2(1 \rm{Al})$) [Fig.~\ref{fig3}(a)]. Dimers form by precipitation processes in the supersaturated pore solution before polymerizing into C-S-H phases with larger chain lengths $Q^2$ and $Q^2(1 \rm{Al})$ [Fig.~\ref{fig3}(b)]. This polymerization process is especially pronounced for low values of $\xi_{\rm NMR}$. However, while the formation rate of C-S-H particles with longer silicate chains seems  to be independent of $\xi_{\rm NMR}$ (compare the slope of $Q^2$ curve for 0.1~$\leq \xi_{\rm NMR} \leq$~0.4 and for $\xi_{\rm NMR} \geq$~0.6), the fraction of $Q^1$ sites decreases slightly for 0.25~$\leq \xi_{\rm NMR} \leq$~0.4 but then undergoes a clear increase for $\xi_{\rm NMR} \simeq 0.5$. The small decrease in the fraction of $Q^1$ sites may reflect that dimers join by a silicate bridging tetrahedron, forming pentamers or longer chains. The subsequent increase in $Q^1$ sites is concomitant to the decrease in average C-S-H chain length (i.e., the average number of silicate tetrahedra, including tetrahedral $Al$, in the chains), as seen in Fig.~\ref{fig3}(b) and is in good agreement with previous results from studies of white Portland cement hydration \cite{Andersen:2003}. This observation may reflect an increased incorporation of Ca$^{2+}$ ions with respect to the number of silica tetrahedra in the interlayer of the C-S-H structure, which will split longer chain units into dimers in the C-S-H, as observed when the Ca/Si ratio of the C-S-H phase increases.

However, despite a clear increase in average chain length (compared to the standard deviation of approximately 0.15) for low degrees of hydration $\xi_{\rm NMR} \leq 0.4$, no correlation between the chain length and the creep modulus $C$ can be observed [Fig.~\ref{fig3}(c)], confirming that early age creep occurs at the C-S-H interparticle-scale and not at the sub-nanometer (molecular) scale as probed by solid state NMR. This provides additional evidence that the creep behaviour of the phases is not controlled by the internal structure or nanoscale properties of the individual C-S-H-particles, but rather happening at a larger spatial scale, for instance  by a sliding between particles. In this context, we emphasize that the drop in chain length observed for $\xi_{\rm NMR}$ values between approximately 0.38 and 0.50, goes along with substantial changes in interparticle spacing $r/d$ [see Fig.~\ref{fig2}(b)]. This indicates that changes in the polymerization of the C-S-H do not influence the creep behaviour directly and that these changes (i.e., polymerization) are related to the formation of new particles thus indirectly influencing the cement paste microstructure.    
The degree of hydration $\xi_{\rm NMR}$ at which the changes in fractions of $Q^1$ and $Q^2$ sites occurs are the same degree of hydration beyond which there is a sudden increase in the fraction of LD-C-S-H phases, shown in Fig.~\ref{fig1}(d). 

\begin{figure}[t!]
\centering
\includegraphics[width=\columnwidth]{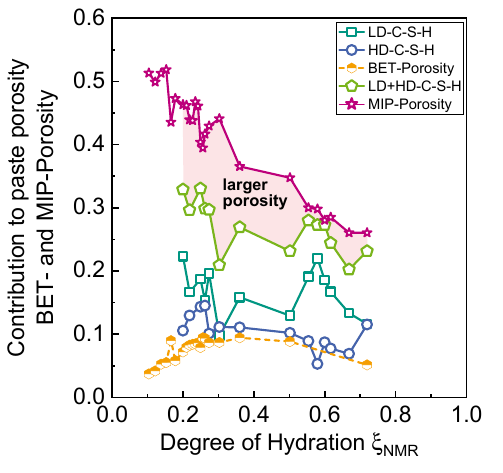}
\caption{(color online) Contribution to the cement paste total porosity of LD-C-S-H, HD-C-S-H and both phases as determined by statistical nanoindentation as a function of the degree of hydration $\xi_{\rm NMR}$. Results are compared to direct porosity measurements performed by nitrogen adsorption at 77K and Mercury Intrusion Porosimetry (MIP).
\label{fig4}}
\end{figure} 

\subsection{Creep response in light of cement's porous microstructure} 
Final indirect confirmation for the scenario described above comes from the investigation of the pore size distribution of cement hydrates. From the mechanical results, we compute the contribution of the LD- and HD-C-S-H phases porosity to the overall porosity of the cement paste at various degrees of hydration. The porosity $p_i$ of the phase $i$ is calculated using $p_i=f_i(1-\eta_i)$, where $f_i$ designates the volume fraction  [Fig.~\ref{fig1}(d)], and $\eta_i$ the packing density of the phase $i$ [Fig.~\ref{fig1}(h)]. As can be seen in Fig.~\ref{fig4} [symbols (\textcolor{mygreen}{\textbf{$\pentagon$}})], the joint contribution of LD- and HD-C-S-H phases to the total porosity of the sample are quasi constant, independent of the degrees of hydration, with a mean value of 0.28, which coincides remarkably well with the value of 0.27 predicted by Powers in his seminal work \cite{Powers:1949}. 
These estimates are compared with direct measurements of the porosity associated with pores of radius smaller than 20~nm and determined by nitrogen adsorption at 77K (see Appendix~D for details). The porosity measured by adsorption as a function of the degree of hydration shows a bell-shaped curve with a maximum at $\xi_{\rm NMR} \simeq 0.4$ [symbols (\textcolor{myyellow}{\hexagofillha}) in Fig.~\ref{fig4}], which points toward a change in the hydration process for high degrees of hydration similar to what has been found in both nanoindentation and \textsuperscript{29}Si NMR testing. This change has been associated with a transition from a free nucleation and growth process of C-S-H particles, to a diffusion controlled process \cite{FitzGerald:1998}. A similar change is visible in the volume fraction of LD- and HD-C-S-H estimated from the mechanical tests at $\xi_{\rm NMR} \simeq 0.4$  [Fig.~\ref{fig1}(d)]. Indeed, for $\xi_{\rm NMR} \lesssim 0.55$, the space provided by the larger pores is sufficient for the formation of both new LD- and HD-C-S-H, which corresponds to a steady increase of the specific surface area. Whereas, for $\xi_{\rm NMR} \gtrsim 0.55$, the structure formation is more and more dominated by compaction processes of the C-S-H nanoparticles, leading to a reduction of HD-C-S-H porosity. Note that this transition also corresponds to the sudden increase $Q^1$ vs.~$Q^2$ fraction, as observed in Fig.~\ref{fig3}.

Finally, the larger scale porosity of the samples was determined using Mercury Intrusion Porosimetry (MIP), which gives access to the porosity in the range between approximately 5~nm and $500~\mu$m \cite{Aligizaki:2006}. The difference between the porosity determined by MIP [symbols (\textcolor{pink}{$\star$}) in Fig.~\ref{fig4}], i.e., the total pore volume, and the pore volume contribution from LD- and HD-C-S-H phases can be interpreted as the capillary pore volume, i.e., the pores greater than approximately 30~nm in radius \cite{Mindess:2003,Powers:1949}. The results detailed in Fig.~\ref{fig4} clearly prove that for increasing degree of hydration new C-S-H-gel is formed in the larger pores, reducing the overall porosity of the sample.
The total volume-fraction increase of LD- and HD-C-SH phases and the corresponding contribution to porosity appear to be compensated by the increase in packing density of those phases. In agreement with the reasoning of Powers based on macroscopic experiments \cite{Powers:1949}, these new microstructural results show that the overall contribution of the C-S-H-gel to the porosity remains roughly constant. This observation is linked to the continuous compaction of the system, the increasing lack of free space for C-S-H formation, and especially to the fact that the C-S-H particles have a distinct size, posing an upper bound for compaction. More detailed research on the exact mechanisms is, however, necessary.

\begin{figure}[t!]
\centering
\includegraphics[width=0.85\columnwidth]{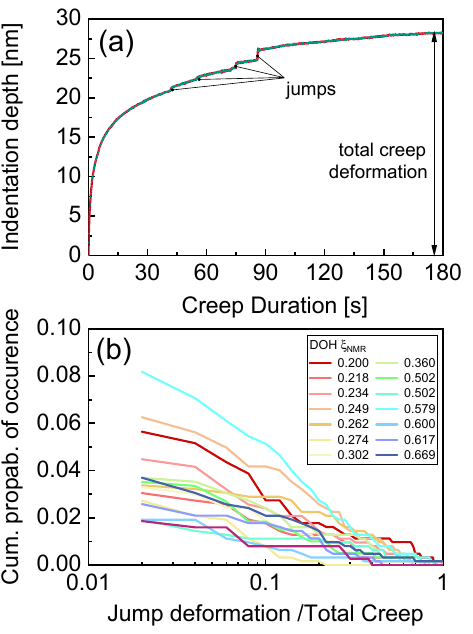}
\caption{(color online) (a) Example of indentation depth vs. time curve afflicted with discontinuities (jumps); (b) Cumulative probability of discontinuities and their contribution to the total creep deformation during the indentation creep phase for the investigated degrees of hydration $\xi_{\rm NMR}$.
\label{fig7}}
\end{figure} 

\subsection{Contribution of crack-formation to the overall creep response} 
\label{crack}
In this section, we discuss the outliers detected in the indentation tests, i.e., the force-depth curves in which the creep phase shows irregularities such as a discontinuous behavior or jumps as illustrated in Fig.~\ref{fig7}(a), and that may be interpreted as a micro-crack. Concerning the question of whether and to which extent creep results from an irreversible microcracking proposed by Wittmann and Bazant \cite{Wittmann:1982,Bazant:1997}, these ``irregular" curves do contain important information. For each creep curve containing jumps, the total jump deformation was summed up from the individual jumps and compared to the total creep deformation (including jumps). Hereby, only jumps with an individual minimum jump height of 1~nm were considered, corresponding to three times the standard deviation of the measurement noise. The cumulative probability of jumps and their contribution to the total creep deformation is reported in Fig.~\ref{fig7}(b) for various degrees of hydration. Only approx.~3\% to 8\% of all indentation creep vs.~time curves showed discontinuities or jumps. In case jumps do occur, they can contribute significantly to the measured creep deformation, however with the likeliness of large jumps to occur decreasing strongly. Moreover, no clear influence of the degree of hydration $\xi_{\rm NMR}$ on the jump height was visible. 

Coming back to the question, whether cement paste creep is significantly controlled by abrupt microcracking processes, our data indicate that this process eventually takes place but is negligible compared to other processes. This conclusion however is subject to the restriction that jumps in depth during the creep phase can also result from external mechanical vibrations. An analysis was carried out investigating the likeliness of a jump to occur as a function of the time of day - and thus of the likeliness of external vibrations in the building. We found that jumps occur both during day and night-time, however  the likeliness of jumps strongly increases during daytime. This finding thus indicates, that the contribution of micro-cracking to the total deformation is even lower than that suggested by Fig.~\ref{fig7}(b).

\begin{figure}[t!]
\centering
\includegraphics[width=\columnwidth]{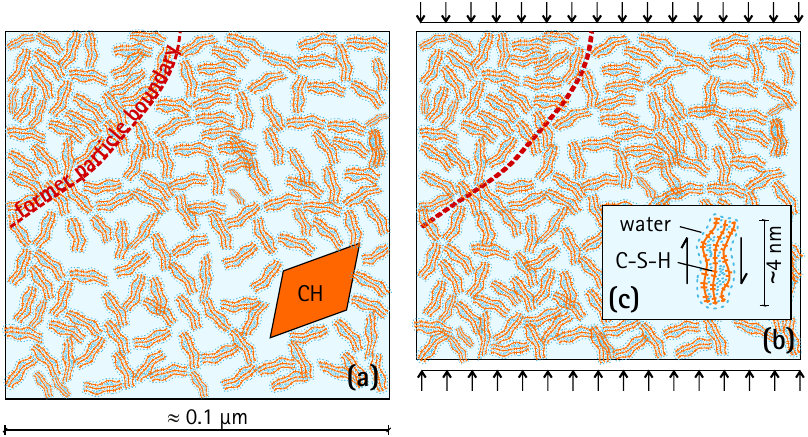}
\caption{(color online) Sketch of the microstructure of hardened cement paste in (a) the unloaded stage and (b) the loaded stage.Different areas of the sample show different packing densities (LD- and HD-C-S-H) associated with the spaces formerly occupied by particles prior to their dissolution (see dashed red line). Deformation of the individual C-S-H particles (see inset (c)) subjected to loading as proven in Molecular Dynamics simulation might facilitate re-organization of densely packed system and thus allow creep.
\label{fig8}}
\end{figure} 

\section{Discussion and Conclusion} 
\label{DiscussionConclusion}
The present study provides strong experimental evidence that the creep response of hardening cement paste is primarily determined by the packing density of the C-S-H nanoparticles that compose the paste microstructure. We have shown that the creep modulus  increases linearly for increasing packing density over the range $0.55\leq\eta \leq 0.74$, or equivalently for decreasing distance between particles. 
This result suggests that the creep response of hardened cement paste mainly results from viscous sliding between C-S-H nanoparticles separated by nanoconfined water layers (see the sketch in Fig.~\ref{fig8}). This microscopic picture of the creep response follows the same line of thought as the seminal considerations by Powers \cite{Powers:1968} and Wittmann \cite{Wittmann:1982}, who suggested that the contribution to the total creep deformation, of a viscous deformation of individual C-S-H particles, is negligible. Moreover, our experimental findings strongly support the conclusions of recent Molecular Dynamics simulations of hydrated cement paste \cite{Morshedifard:2018}, in which the authors attribute the logarithmic creep response of cement paste (similar to the one we observe) to the sliding of C-S-H particles, provided that the water content is sufficient.
In that context, our results allow ruling out the scenario following which creep in HCP results from a microcracking of the paste. 

Finally, our results show that, in the limit of high degrees of hydration, the packing density of C-S-H particles has reached its maximum value. Such a jammed state of the hardened paste microstructure should either hinder creep deformation or lead to a dilation of the system under external stress. Yet, such effects are not observed. Therefore, for strongly hydrated system, our results suggest that the viscoelastic responses of individual C-S-H particles eventually come into play to facilitate the inter-particle sliding \cite{Cates:1998,McDowell:2003}, and allow for the material to creep. This picture is compatible with numerical results reported in refs. \cite{Manzano:2013,Morshedifard:2018}, in which C-S-H particles themselves can be subject to a creep deformation, which strongly facilitates the yielding process. 

In conclusion, the combination of nano-mechanical testing together with $^{29}$Si NMR, and gas adsorption and mercury intrusion testing provides a comprehensive microscopic picture of the creep response of hydrating cement pastes over a large range of hydration degrees. Our results allow to rank the three scenarios historically proposed to account for the creep deformation in cement paste. Beyond hydrating cement pastes, future work will help determining the extent of our results to other reactive granular materials.  

\begin{acknowledgments}
This work was funded by a research stipend of the Deutsche Forschungsgemeinschaft (DFG) to Michael Haist (reference HA 7917/1-1). The authors thank N. Chatterjee (MIT-EAS) for his support in carrying out the WDS-mappings as well as K.~Ioannidou, S.~Yip, P.~Stemmermann, K.~Garbev and T.~Petersen for extremely fruitful discussions.
\end{acknowledgments}

\appendix
\section{Sample preparation} 
All cement paste samples were prepared by mixing ordinary Portland cement CEM I 42.5 R (65\% wt. tricalcium-silicate Ca$_3$SiO$_5$; 15\% wt. dicalcium-silicate Ca$_2$SiO$_4$; 5\% wt. tricalcium-aluminate Ca$_3$Al$_2$O$_6$; 2\% wt. tetracalcium-aluminate-ferrite Ca$_4$Al$_2$Fe$_2$O$_{10}$)  produced by Wittekind (Germany) with demineralized water, at a water to cement mass ratio  $w/c=0.4$. Samples were cast in cylindrical tubes of length 50~mm and diameter 24~mm before being stored at 20$^{\circ}$C in a saturated lime (CaO) solution. Following \cite{Zhang:2011}, the hydration reaction was stopped by solvent exchange with isopropanol at various points in time ranging from 7 hours to 220 days after water addition to the dry cement. Once the hydration stopped, the cylindrical sample were then cut into discs, which surfaces were polished manually by using a series of SiC papers so as to reach a surface roughness of a few hundred nanometers \cite{Miller:2008}. Finally, the polished samples of hardened cement paste were dried in a thermal chamber at 60$^{\circ}$C until they reached a constant mass. They were then tested mechanically by nanoindentation.\vspace{-0.2cm}
 
\section{Statistical nanoindentation} 
The mechanical properties of the cement samples were characterized by nanoindentation (UNHT, Anton Paar). For each sample, $25\times25=625$ indents separated by 10$\mu$m were carried out on a square map using a three-sided, pyramid-like Berkovich indenter. Each indentation test consists in measuring the indentation depth of the indenter resulting from a symmetric load profile: ramp of increasing load up to a constant value $P_{\rm max}$ maintained for 180s so as to determine the creep response of the sample, followed by a ramp of decreasing load at a rate identical to the loading part. The maximum load $P_{\rm max}$ is chosen to yield indentation depths of approximately 300~nm, probing the mechanical local properties of the hardened cement paste within a volume of about 1~$\mu$m$^{3}$. For each indent, the local values of the hardness $H$ and the indentation modulus $M$ were determined by the method of Oliver and Pharr \cite{Oliver:2004}, while the creep modulus $C$, which corresponds to the inverse of a creep rate, was determined following the method previously used by Vandamme and Ulm in \cite{Vandamme:2009,Vandamme:2013} (also see Sec.~\ref{results}). For a given sample, an indentation grid yields a set of 625 triplets ($H$, $M$, $C$), which is analyzed with a Gaussian Mixture Modeling extensively described in \cite{Krakowiak:2015}. Such analysis, combined with a Bayesian information criteria, allows us to determine the most likely number of phases based on the Gaussian clustering of the mechanical properties and the chemical content at the locus of each indent (see following paragraph). The analysis yields between 3 and 5 phases, which correspond to the phases composing the hardened cement paste: low-density C-S-H (LD-C-S-H), high-density C-S-H (HD-C-S-H), Calcium hydroxide (CH), mixed phases and unhydrated clinker. \vspace{-0.2cm}

\section{Chemical surface mapping}  
The entire surface of the hardened cement paste sample is coated with a carbon layer of 20~nm thickness and dried in ultra-vacuum ($8\cdot10^{-9}$bar) for at least 24 hours. The chemical composition of the sample at the locus of the indentation grid is determined using Wavelength-Dispersive X-ray Spectroscopy (WDS) with a scanning electron microscope (SEM, JEOL JXA-8200). SEM backscatter images and WDS maps with dimensions of 368~$\mu$m$\times$276$~\mu$m (corresponding to 1024$\times$768~pixels) were acquired with a resolution of 0.36$\mu$m. The beam voltage, current, working distance and dwell time were set respectively to 15~kV, 10~nA, 11~mm and 40~ms per spot. The WDS mapping was performed for the following species: calcium (Ca), silicon (Si), aluminum (Al), iron (Fe), sulfur (S), magnesium (Mg), sodium (Na) and potassium (K).\vspace{-0.2cm}

\section{Bulk testing techniques} 
Single-pulse \textsuperscript{29}Si MAS NMR spectra were acquired on a Bruker 400~MHz (9.39~T) spectrometer using a home-built CP/MAS probe for 7~mm outer diameter rotors, a 45 degree excitation pulse, a 30~s relaxation delay and typically 2048 scans. The resulting spectra were deconvolved providing relative intensities for alite, belite, and the $Q^1$; $Q^2$, and $Q^2$(1Al) resonances of the C-S-H phase. From these intensities, the degrees of silicate hydration, $\xi_{\rm NMR}$ = [$I(Q^1) + I(Q^2) + I(Q^2(1Al))$]/$I_{tot}$ (with $I_{tot}$ designating the total signal intensity) and the average chain length of alumino-silicate tetrahedra, CL = 2[$Q^1 + Q^2 + 3/2Q^2(1Al)]/Q^1$, were calculated (see \cite{Andersen:2003}). Moreover the porosity of the hardened cement paste was determined by Mercury Intrusion Porosimetry MIP using a Micromeritics Autopore III 9420 instrument with Hg pressures up to 400~MPa and by Nitrogen adsorption at 77~K using a Micromeritics ASAP 2000 instrument. Here the MIP method was used to measure the total porosity of the sample, with pores as small as approximately 5~nm in radius \cite{Aligizaki:2006}. The pore-size-distribution accessed by Nitrogen adsorption was derived using the Barrett-Joyner-Halenda (BJH) algorithm, covering pores with a pore diameter between 1~nm and approx.~120~nm, thus encompassing both LD-, HD-C-S-H and parts of the capillary porosity (i.e., pores with a radius larger than 20~nm) \cite{Aligizaki:2006}. In order to identify the porosity and pore size distribution of the individual phases identified via the nanoindentation coupled to surface chemistry, a superposition of Gaussian functions was fitted to the pore volume vs. log pore radius distribution data obtained in BET testing thus identifying the pore size clusters present in the paste. At least 3 clusters could be identified with mean pore radii of 6~nm, 15~nm and, 28~nm. Hereby, the identified mean pore radii of the first two clusters (i.e., 6~nm and 15~nm) correspond well to the sizes derived from NMR relaxometry data on similar pastes (see \cite{Halperin:1994}). The pore volume contribution of the first two clusters as shown in Fig.~\ref{fig4} (designation BET-Porosity) was calculated by integrating the pore size distribution in the range between 1~nm to 20~nm, where the upper threshold results from the mean pore radius of cluster 2 (i.e., 15~nm) plus 3 times its standard deviation. These results suggest that the BET pore clusters no.~1 and 2 (i.e., 6~nm and 15~nm) can be attributed to HD-C-S-H, whereas the properties of LD-C-S-H are significantly affected by larger pores. It should be noted, that this finding is somewhat in contradiction of rough estimates on the pore radius of LD-C-S-H by Jennings \cite{Jennings:2000}, which however were not measured but calculated from estimates of density and specific surface and thus are prone to various possible errors.


\begin{thebibliography}{73}
\bibitem [{not()}]{note1}%
  \BibitemOpen
  \href@noop {} {}\bibinfo {note} {The influence of aluminate phases could be
  neglected due to the low aluminate content of the investigated
  cement.}\BibitemShut {Stop}%
\end{thebibliography}

%

\end{document}